\documentclass[aps,prl,twocolumn,groupedaddress,showpacs]{revtex4}
\pdfoutput=1

\usepackage{graphicx}
\usepackage{color}
\usepackage{hyperref}

\begin{document}

\title{Low-lying resonances and relativistic screening in Big Bang Nucleosynthesis}

\author{M.A. Famiano$^{1,3}$}
\email{michael.famiano@wmich.edu} 
\author{A.B. Balantekin$^{2,3}$}
\email{baha@physics.wisc.edu}
\author{T. Kajino$^{3,4}$}
\email{kajino@nao.ac.jp}
\affiliation{$^1$Department of Physics, Western Michigan University, Kalamazoo, Michigan 49008, USA}
\affiliation{
$^2$Department of Physics, University of Wisconsin, Madison, WI 53706, USA}
\affiliation{
$^3$National Astronomical Observatory of Japan 2-21-1 
Osawa, Mitaka, Tokyo, 181-8588, Japan}
\affiliation{
$^4$Department of Astronomy, School of Science, 
The University of Tokyo, 7-3-1 Hongo, Bunkyo-ku, Tokyo, 113-0033, Japan}

\begin{abstract}
We explore effects of the screening due to the relativistic electron-positron plasma and presence of resonances in the secondary reactions leading to $A=7$ nuclei during the Big Bang Nucleosynthesis. In particular, we investigate and examine possible low-lying resonances in the $^7$Be($^3$He, $\gamma$)$^{10}$C reaction and examine the resultant destruction of $^7$Be for various resonance locations and strengths.
While a resonance in the $^{10}$C compound nucleus is thought to have negligible effects we explore the possibility of an enhancement from plasma screening that may adjust the final $^7$Be abundance.
We find the effects of relativistic screening and possible low-lying resonances to be relatively small in the standard Early Universe models. 
\end{abstract}


\maketitle 

\section{Introduction}

Observation of the accelerated expansion of the Universe, measurement of the Cosmic Microwave Background Radiation temperature anisotropies and the observation of the light elements produced during the Big Bang Nucleosynthesis (BBN) epitomize the current status of the precision cosmology. 
Especially the BBN is an ideal tool not only to test aspects of the Standard Models of cosmology as well as of nuclear and particle physics but also to look for new physics beyond those standard models. 
(For recent reviews of BBN see Refs. \cite{Cyburt:2015mya,Coc:2014oia,Iocco:2008va,Steigman:2007xt}). In particular new calculations of the light-element abundances are performed \cite{Cyburt:2015mya} using the recent high-precision 2015 {\em Planck} measurement of the baryon-to-photon ratio, helium abundance and the effective number of relativistic degrees of freedom, $N_{\rm eff}$ \cite{Ade:2015xua}, as well as astronomical observations of deuterium \cite{Burles:1997ez,Cooke:2013cba}. 
These calculations find that D/H observations are now more precise than the corresponding theoretical predictions, but predictions for $A=7$ nuclei continue to disagree with observations. 
A recent update (NACRE II) of the compilation of charged-particle-induced thermonuclear reaction rates for nuclei with mass number $A<16$ \cite{Xu:2013fha} was used in the calculations of Refs. \cite{Cyburt:2015mya} and \cite{Coc:2014oia}. These calculations also provide confidence limits for the production of $^6$Li, $^9$Be, $^{11}$B, and carbon, oxygen, and nitrogen (CNO). A precise value of the CNO/H is not only important for population III star formation, but as we elaborate in this paper, may also affect the abundances of $A=7$ nuclei. 

The purpose of the present paper is twofold. One is to explore the plasma effects in the BBN more completely. It was already demonstrated that non-relativistic screening of the Coulomb interaction in nuclear reactions taking place during the BBN epoch does not produce a noticeable impact on light element abundances \cite{Wang:2010px}. However, the screening effects due to the relativistic electron-positron plasma was not included in the analysis of Ref. \cite{Wang:2010px}. It was recently shown that electron-positron plasma screening is crucial for neutrino interactions in the BBN epoch 
\cite{Vassh:2015yza}. In the current work we include the effects of the screening due to the electron-positron plasma in nuclear reactions during the BBN epoch. A second purpose is to explore the ramifications of the possible presence of resonances in relatively unexplored reactions involving $A=7$ nuclei. 

\section{Effects from Screening in a Relativistic Electron Plasma}

\begin{figure}
	\includegraphics[width=\linewidth]{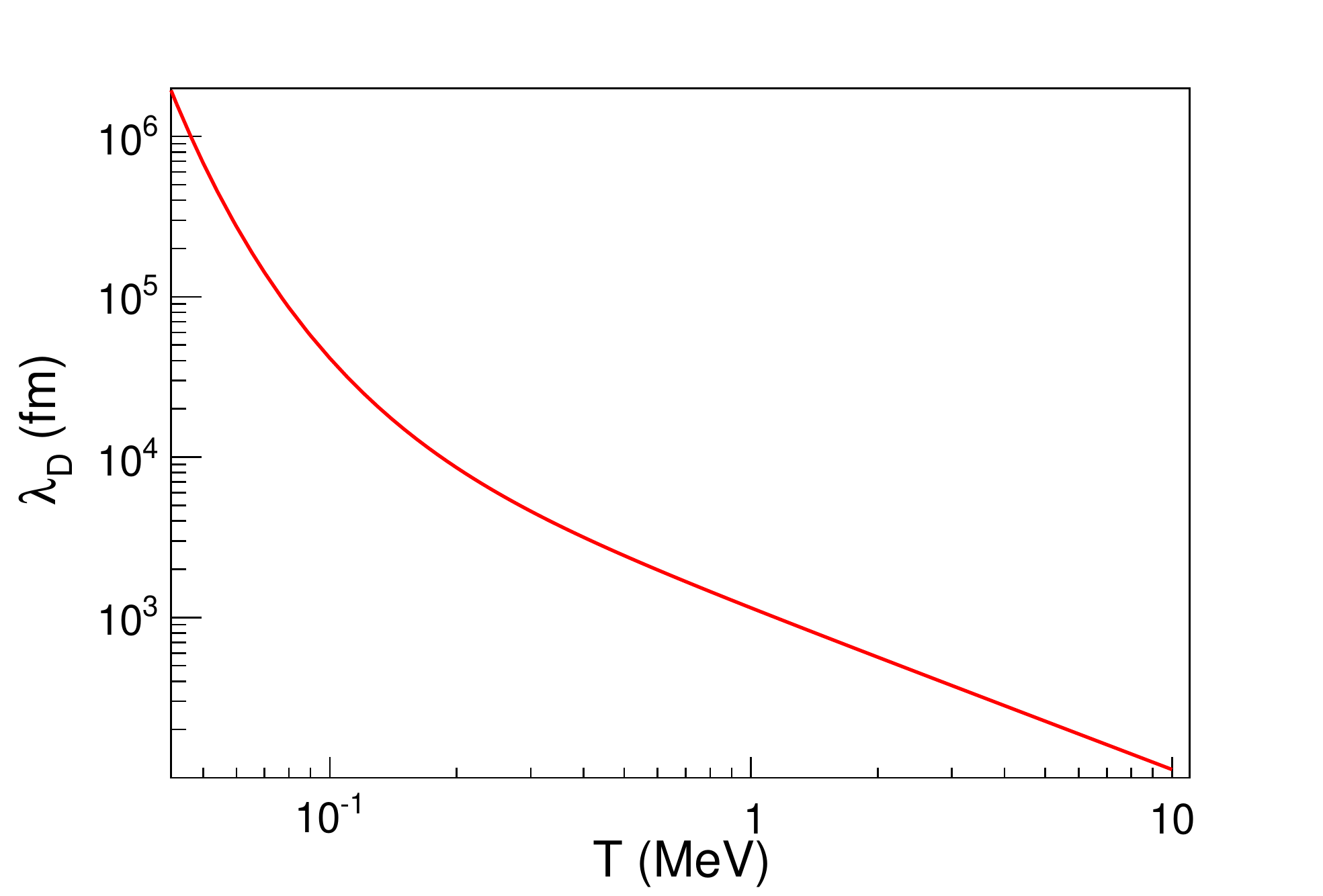}
	\caption{\label{debye_zeromu}The Debye length for a relativistic electron plasma as a function of temperature at $\mu=0$.}
\end{figure}

For a low-density plasma at T$\sim$1 to 2 MeV,  non-relativistic screening can be 
neglected as the associated Debye length is $\lambda_D\sim10^4$ fm (see Appendix) \cite{Wang:2010px}. For example, for $Z_1 = Z_2 =2$, and temperature of 1 MeV, one gets  $\lambda_D =10^4$ fm and the Salpeter correction to the reaction rate $f_D-1 \approx 6 \times 10^{-4}$. However, for an electron-positron plasma, as would exist in the BBN epoch, the temperature is of the same order as the electron mass, requiring the relativistic expression (given in the Eq. (\ref{a3}) of Appendix).  For a vanishing chemical potential, this Debye length as a function of temperature is shown in Figure \ref{debye_zeromu}.
At higher temperatures, this is much smaller than the Debye length which was determined for screening by non-relativistic  electrons alone.

\begin{figure}
\includegraphics[width=\linewidth]{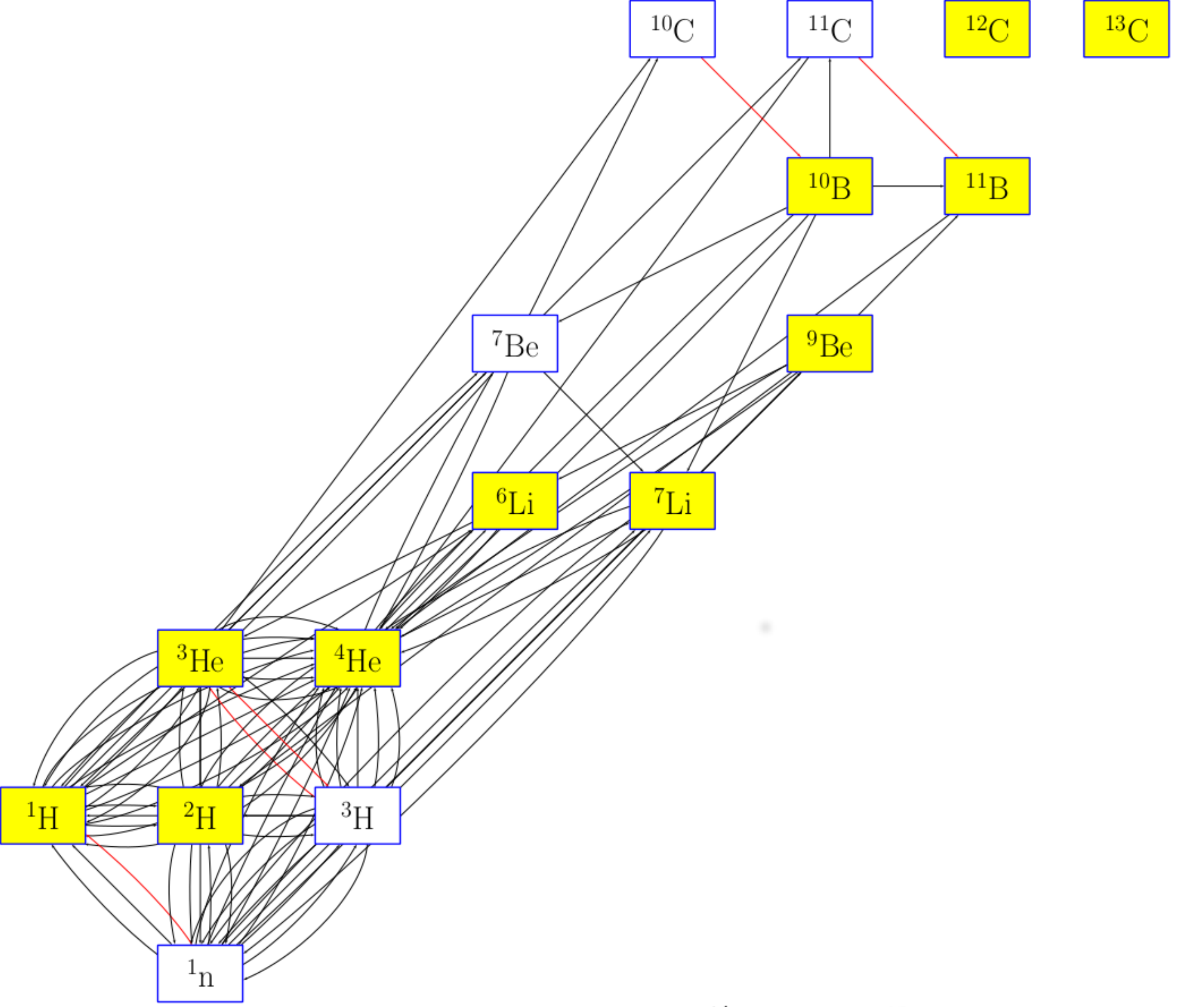}
	\caption{\label{network}The BBN reaction network used up to and including nuclei $Z=6$.}
\end{figure}

Using this Debye length, we used a BBN nuclear reaction network to determine the change in mass fractions based on the
screening enhancement factor.  Corrections were made to reaction rates in the network by inserting the Debye length for a relativistic plasma 
into Eq. (\ref{a5}). The reaction network used is shown in Figure \ref{network}. Reactions up to and including $Z=6$ were included, as well as weak rates and decays. 

\begin{figure}
	\includegraphics[width=\linewidth]{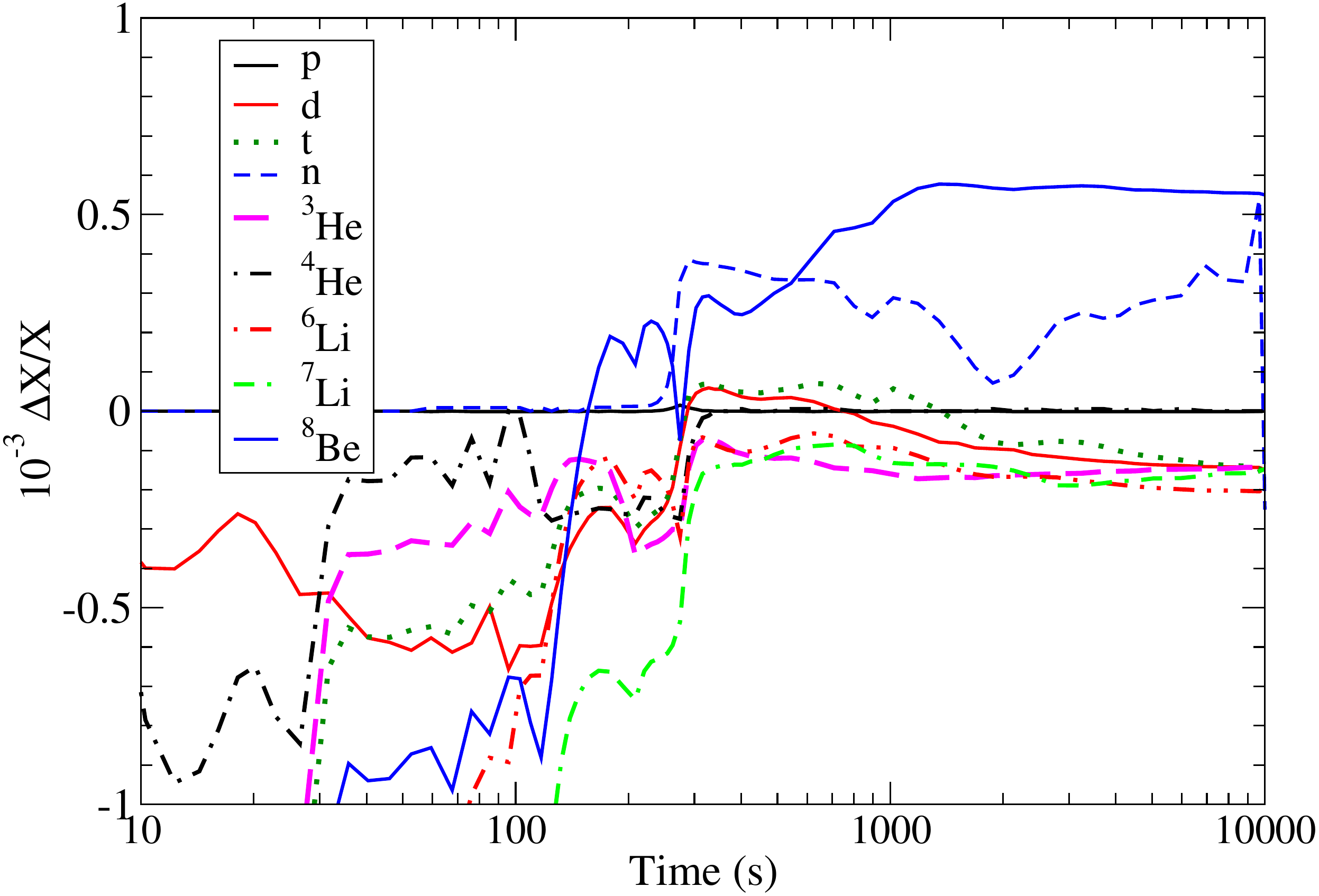}
	\caption{\label{delX_zero}Relative change in mass fractions for nuclei
		in the BBN network as a function of time.}		
		\end{figure}

Mass fractions $X_{\rm bare}$ were determined using a network with unscreened rates.  These are compared to mass fractions $X_{\rm scr}$ from a network employing rates from reactions screened by a relativistic electron plasma.  The change in mass fraction:
\begin{equation}
\frac{\Delta X}{X}\equiv\frac{X_{\rm scr}-X_{\rm bare}}{X_{\rm bare}}
\end{equation}
is shown in Figure \ref{delX_zero}. At early times in the network, the temperature is higher, and the consumption of 
protons, deuterium, and helium proceeds at a higher rate for the
screened reactions compared to the unscreened case. The overall net destruction of these lighter elements is higher
in the screened case.  While there is an overall reduction
in some of the heavier elements, the relative change for those is 
extremely small. Reaction screening is likely to be small in the regime of zero electron chemical potential.  
\subsection{Non-Zero Chemical Potential} 
Inclusion of a non-zero chemical potential in Equation \ref{a3} will increase the inverse screening length and
thus decrease the Debye length, thus increasing the reaction rate enhancement factor.

\begin{figure}
	\includegraphics[width=\linewidth]{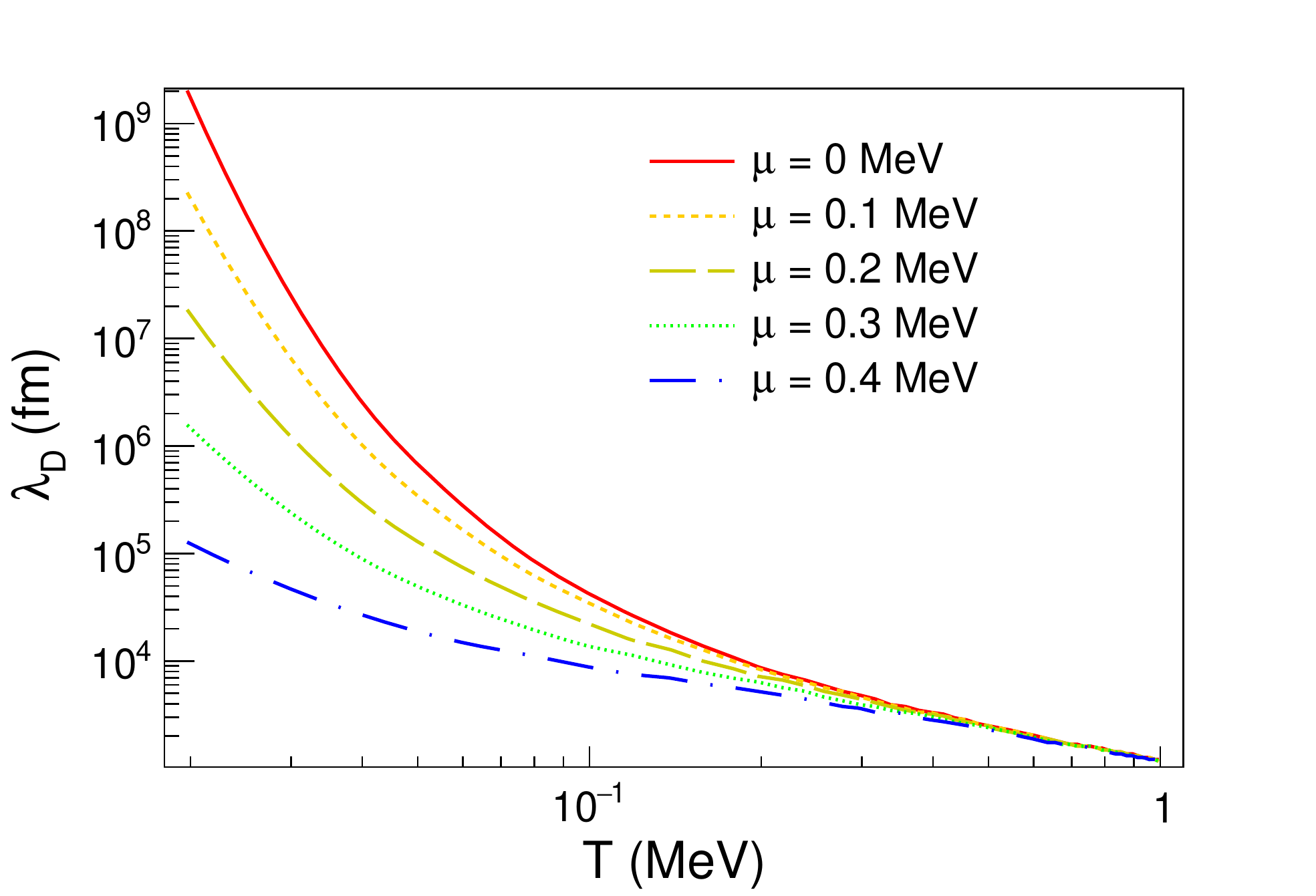}
	\caption{\label{debye_nonzero}Debye length as a function of temperature in an electron plasma for non-zero chemical potential for various chemical
		potentials.}
\end{figure}

\begin{figure}
	\includegraphics[width=\linewidth]{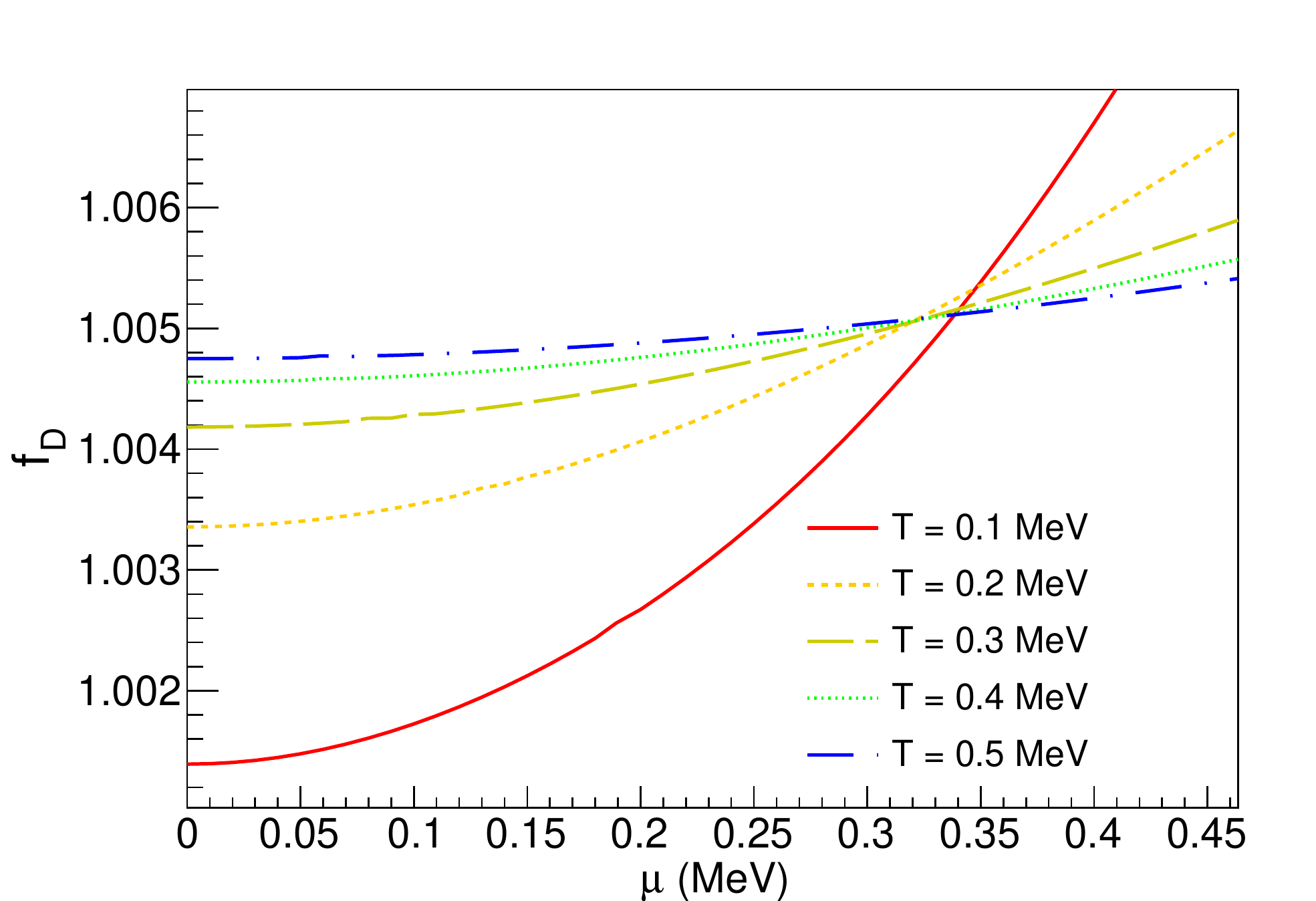}
	\caption{\label{f_mu}Reaction rate enhancement factor as a function of the electron chemical potential for temperatures $T=0.1$ through $0.5$ MeV with $Z_1=Z_2=2$.}
	\end{figure}

For a non-zero chemical potential, Equation \ref{a3} was solved numerically. The Debye length, $\lambda_D$, for various chemical potentials, $\mu$, as a function of temperature is shown in Figure \ref{debye_nonzero}.  A somewhat large chemical potential is necessary for a reduction in the Debye length by one order of magnitude, and this occurs only at low temperature.  At higher temperatures, the electron kinetic energy dominates over the chemical 
potential, and the effect of $\mu$ is reduced. 
The resulting enhancement factor, $f_D$, is shown in Figure \ref{f_mu} for $Z_1=Z_2=2$.  From this figure, one sees that for zero chemical potential the enhancement factor changes more rapidly at low temperature than at high temperature.  As the temperature approaches the value of the electron mass, the enhancement factor increases less rapidly. One also sees that the enhancement factor as a function of $\mu$ changes much less
at higher temperatures than at lower temperatures.  

\begin{figure}
	\includegraphics[width=\linewidth]{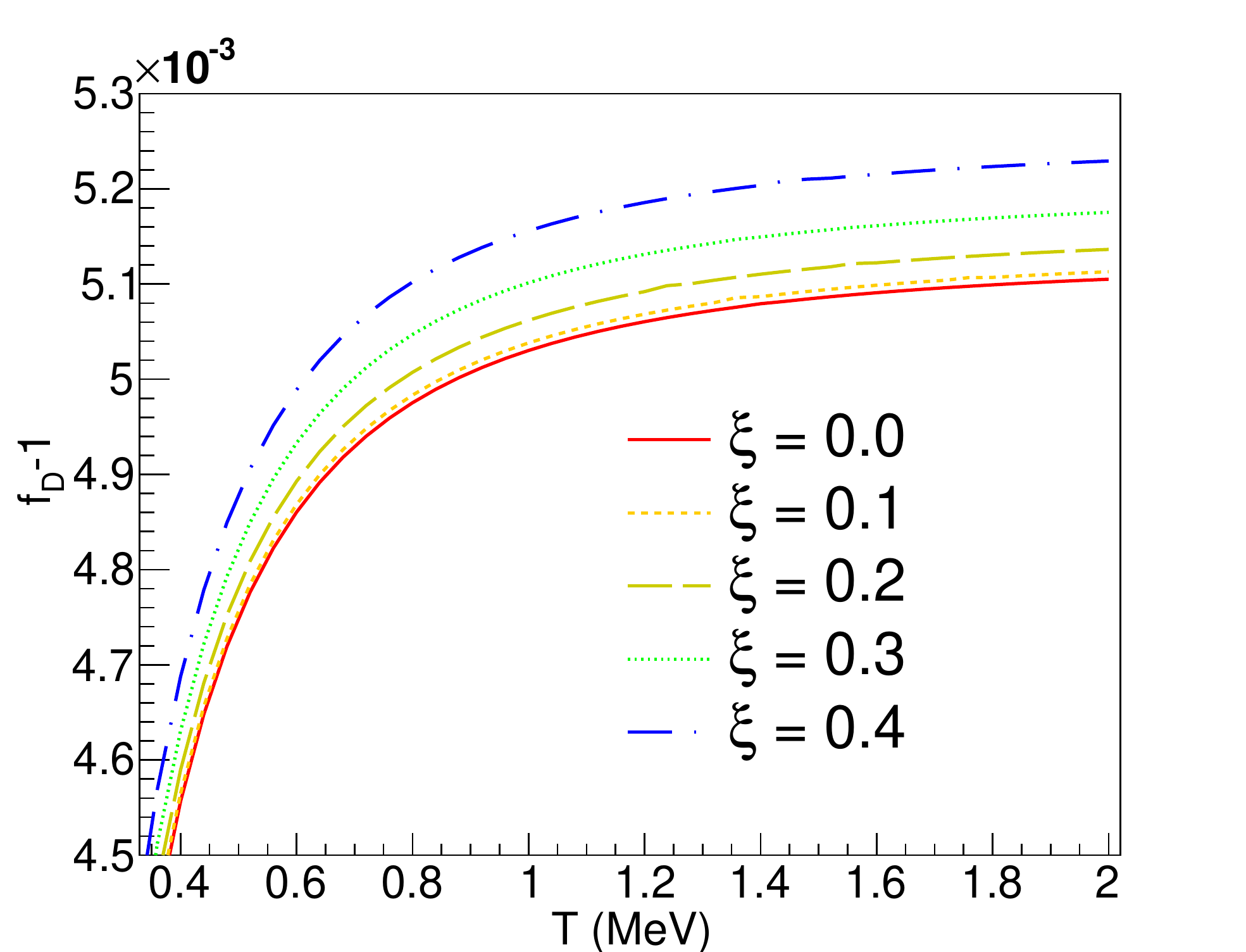}
	\caption{\label{enhancement_zeta}The reaction rate enhancement factor
	as a function of temperature for various electron degeneracy factors with $Z_1=Z_2=2$. In this figure, $f=f_D(\mu \neq 0)$.}
\end{figure}

It is customary to relate the chemical potential to the electron degeneracy parameter:
\begin{equation}
\xi=\frac{\mu}{T} .
\end{equation}
Assuming a constant degeneracy parameter, then the chemical potential is much lower at lower T, reducing the effect of $\mu$ at lower T.  This is shown in Figure \ref{enhancement_zeta}.  One sees that, because the chemical potential now is a linear function of temperature, there is little change in the enhancement factor with degeneracy.  

It is thus concluded that, since the electron degeneracy is very small,  screening from the relativistic electron-positron plasma has little effect on the final abundance distribution in
the standard BBN.

\section{Reduction of $^7$Be From Reactions on $^3$He}
There is significant discrepancy between the observations and the BBN predictions for $A=7$ nuclei, known as the ``lithium problem". (For an overview of the current status of the lithium problem see Ref. 
\cite{Fields:2011zzb}). One alternative mechanism for the possible reduction of $^7$Li was proposed in 
Ref.  \cite{Chakraborty:2010zj}, namely that the consumption of $^7$Be (and subsequently $^7$Li) may
occur through a resonant reaction through the $^{10}$C 
compound nucleus:
\begin{eqnarray}
^7{\rm Be}+^3{\rm He} \rightarrow ^{10}{\rm C}^*
~& \rightarrow  ^{10}{\rm C} + \gamma \\
~& \rightarrow  ^{9}{\rm B} + p \\
~& \rightarrow  2\alpha + 2p  
\end{eqnarray} 

While this reaction has certainly been previously explored \cite{Hammache:2013jdw}, here it is investigated in light of possible resonance structure in the mirror product $^{10}$Be \cite{Yan:2002bc}, particularly at resonant energies greater
than 0.5 MeV. Prior studies have not found resonances in the $^{10}$C 
nucleus \cite{Hammache:2013jdw} for relatively large widths in the range $E_R<500$ keV.  This work 
examines resonances both within and outside 
the energy range and widths previously investigated where the effects of a relativistic electron-positron plasma on 
BBN reaction rates are included. We examine the possible effects of shifts in the thermonuclear reacstion rates for resonant and non-resonant 
reactions from a relativistic electron plasma. 
\subsection{Resonances in $^7$Be+$^3$He}
Taking the resonance structure of $^{10}$Be as motivation, a resonance is assumed for the $^7$Be+$^3$He reaction.  
We note that this reaction and any possible resonances
within this reaction have been shown to have a negligible 
effect on BBN \cite{Broggini12}.  Likewise, there is no experimental evidence for resonances below 500 keV \cite{Hammache:2013jdw}.  
The effects of screening in the hot BBN plasma from
electrons and positrons has been investigated neither
on the non-resonant nor resonant rates. 
Any possible resonances at E$_R>500$ keV and their corresponding strengths are not known. 
The decay width $\Gamma$ of the $^{10}$Be mirror nucleus for the
17.79 MeV state (0.53 MeV above the reaction threshold) is
known to be $110 \pm 35$ keV \cite{Ajzenberg-Selove:1984nja} though the entrance 
channel width has not been experimentally determined.

We assume a resonance cross section of the form 
\begin{equation}
\label{resonanceXsection}
\sigma (E) = \pi \lambdabar^2 \omega \gamma \frac{\Gamma_{\rm total}}{(E-E_R)^2 + \Gamma_{\rm total}^2/4}
\end{equation}
where 
\begin{equation}
\Gamma_{\rm total} =\Gamma_{\rm in} + \Gamma_{\rm out}, 
\end{equation}
$\gamma$ is the reduced width 
\begin{equation}
\gamma = \frac{\Gamma_{\rm in} \Gamma_{\rm out}}{\Gamma_{\rm total} }, 
\end{equation}
$\lambdabar$ is the de Broglie wavelength, and $\omega$ is the appropriate spin factor. 

In the current study, final BBN mass fractions are determined for reaction rates based on a single resonance E$_R$ of arbitrary strength $\omega\gamma$.  The mass fractions are determined as a function of these two parameters. 

The partial width for the entrance channel
is derived from 
a solution to the spherical wave equation at the nuclear potential radius, and the
functional form 
is~\cite{iliadis}:
\begin{equation}
\Gamma_p(E)=2P_L(E)\gamma_L(a)^2
\end{equation}
where the factor $\gamma_L$ is the reduced particle width at a radius $a$ and is
given in the Wigner limit as:
\begin{equation}
\gamma_L^2(a)=\theta^2_L(a)\gamma_W^2(a)=\theta_L^2\frac{3\hbar^2}{2m_{12} a^2}
\end{equation}
and the penetrability factor $P_L$ is given by the regular ($F_L$) and irregular ($G_L$) Coulomb
functions:
\begin{equation}
P_L(a)=\frac{ka}{F_L^2(ka)+G_L^2(ka)}=\frac{\rho}{F_L^2(\eta,\rho)+G_L^2(\eta,\rho)}
\end{equation}
where $\eta$ is the Sommerfeld parameter and $\rho\equiv ka$.  These are evaluated
numerically at the nuclear
radius.  
In this case, only the $L=0$ terms are used, and the single-particle width
$\theta_L^2=0.5$ is assumed. 

\begin{figure}
	\includegraphics[width=\linewidth]{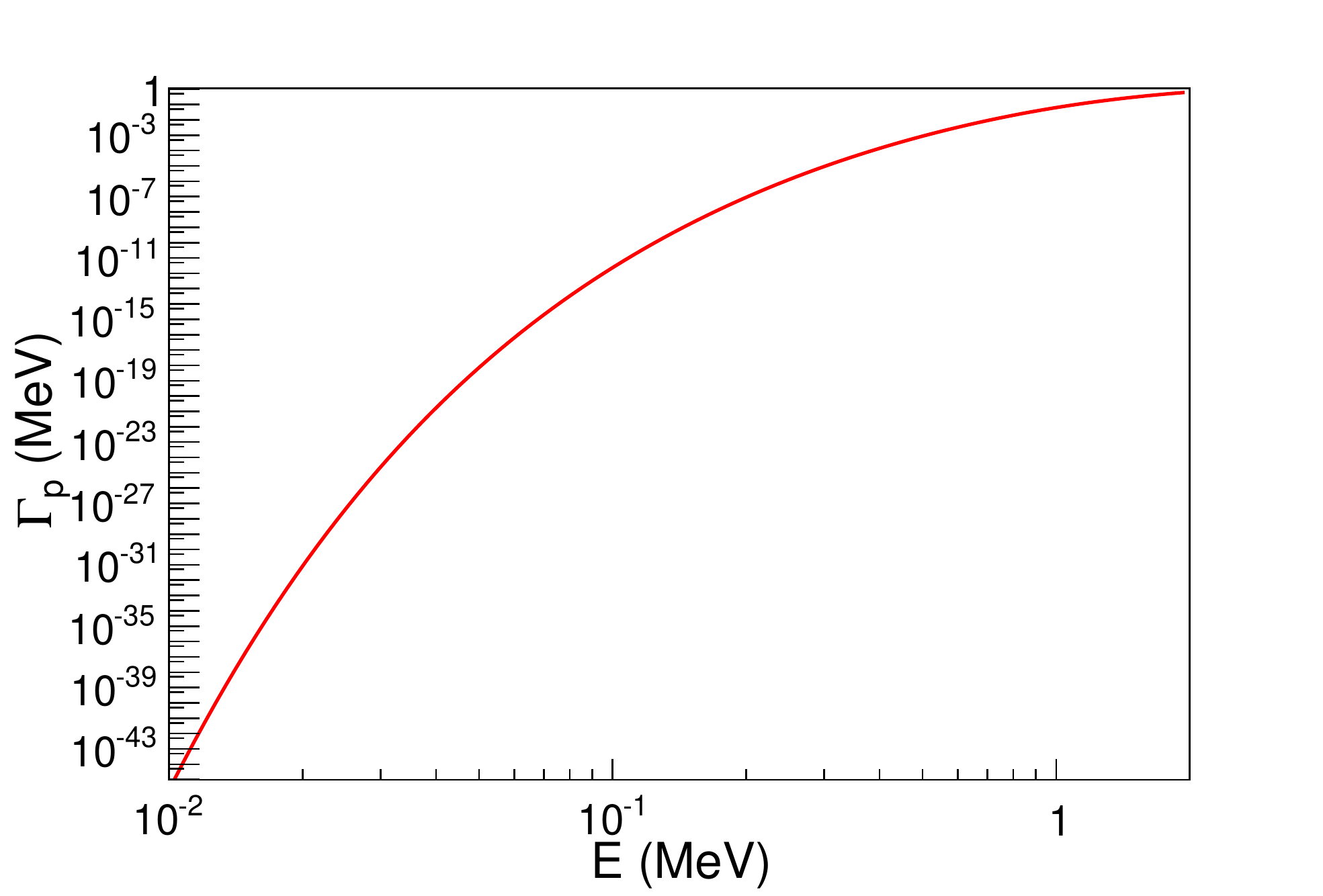}
	\caption{\label{width_e}The assumed entrance channel width as a function of
		incident particle energy for 
		the $^7$Be($^3$He,$\gamma$)$^{10}$C reaction.}
\end{figure}

\begin{figure}
	\includegraphics[width=\linewidth]{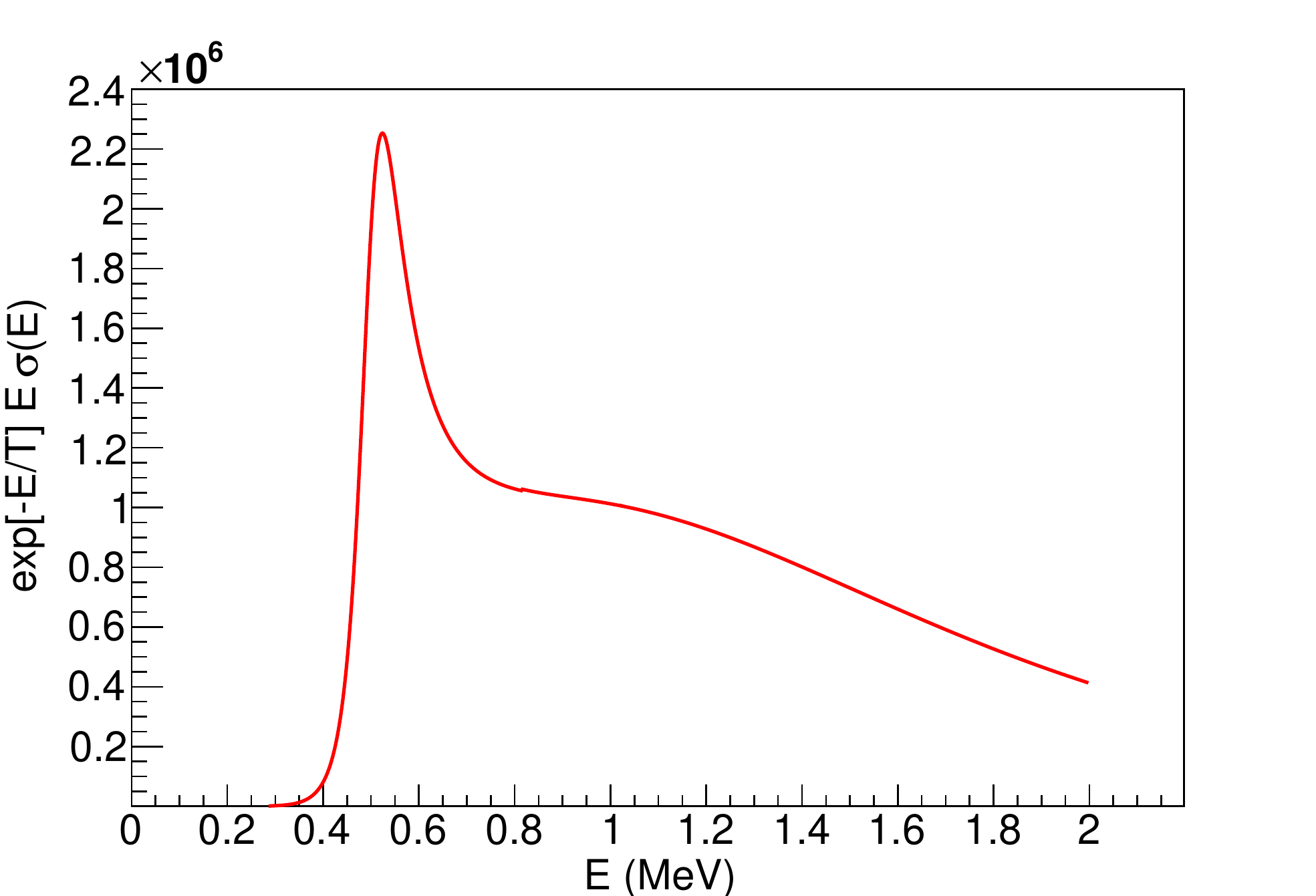}
	\caption{\label{e_dep_sigma}The integrand of the thermonuclear reaction rate for
		the energy-dependent width of
		the $^7$Be($^3$He,$\gamma$)$^{10}$C reaction at T=1 MeV for a resonance located at
		E$_R$=500 keV.}
\end{figure}

The entrance channel width at lower energies is shown in Figure
\ref{width_e} for the $^7$Be+$^3$He reaction.  Near threshold,
the particle width drops 
precipitously, while above threshold, it increases to a substantial value.  
The effect of the energy-dependent width can be shown in Figure \ref{e_dep_sigma}
which shows the integrand of
the thermonuclear rate at $T=1$ MeV for a resonance $E_R=500$ keV.  The high energy
tail of the resonance can
have a large effect on the overall rates. 

Relativistic electron-positron screening can adjust the incident particle energy, effectively 
shifting the threshold energy in the cross-section.  One can see that for near-threshold or sub-threshold resonances, this shift could result in a significant chance in the cross-section,
as discussed in the next section.

In addition to the above resonance structure, we determined BBN abundance distributions for a range of narrow resonances
and strengths.  
By scanning across strengths and resonance locations, the relative reduction of $^7$Be was mapped. 
The mapping of the relative mass fraction of $^7$Be, R, is defined to be the final $^7$Be mass fraction for a  BBN calculation
with a $^{10}$C resonance divided by that with no resonance:
\begin{equation}
R\equiv \frac{X_{res}}{X_{nr}}
\end{equation}
\section{Relativistic Electron Screening and the $^7$Be+$^3$He Reaction}
Prior to examining resonances in the $^7$Be+$^3$He reaction, we proceed with a discussion of the
effects of relativistic electron screening on the same reaction.

While a resonance in $^{10}$C may increase the destruction of $^7$Be, the effect may 
be magnified by the inclusion of screening from the electron plasma.  The enhancement on the cross section is described in the Appendix.
Incorporating screening into the usual thermonuclear reaction rate (TRR) will create an energy shift $E_0$ of the reaction system because of the reduced particle potential.  The energy shift $E_0$ is defined in the appendix.

\begin{figure}
	\includegraphics[width=\linewidth]{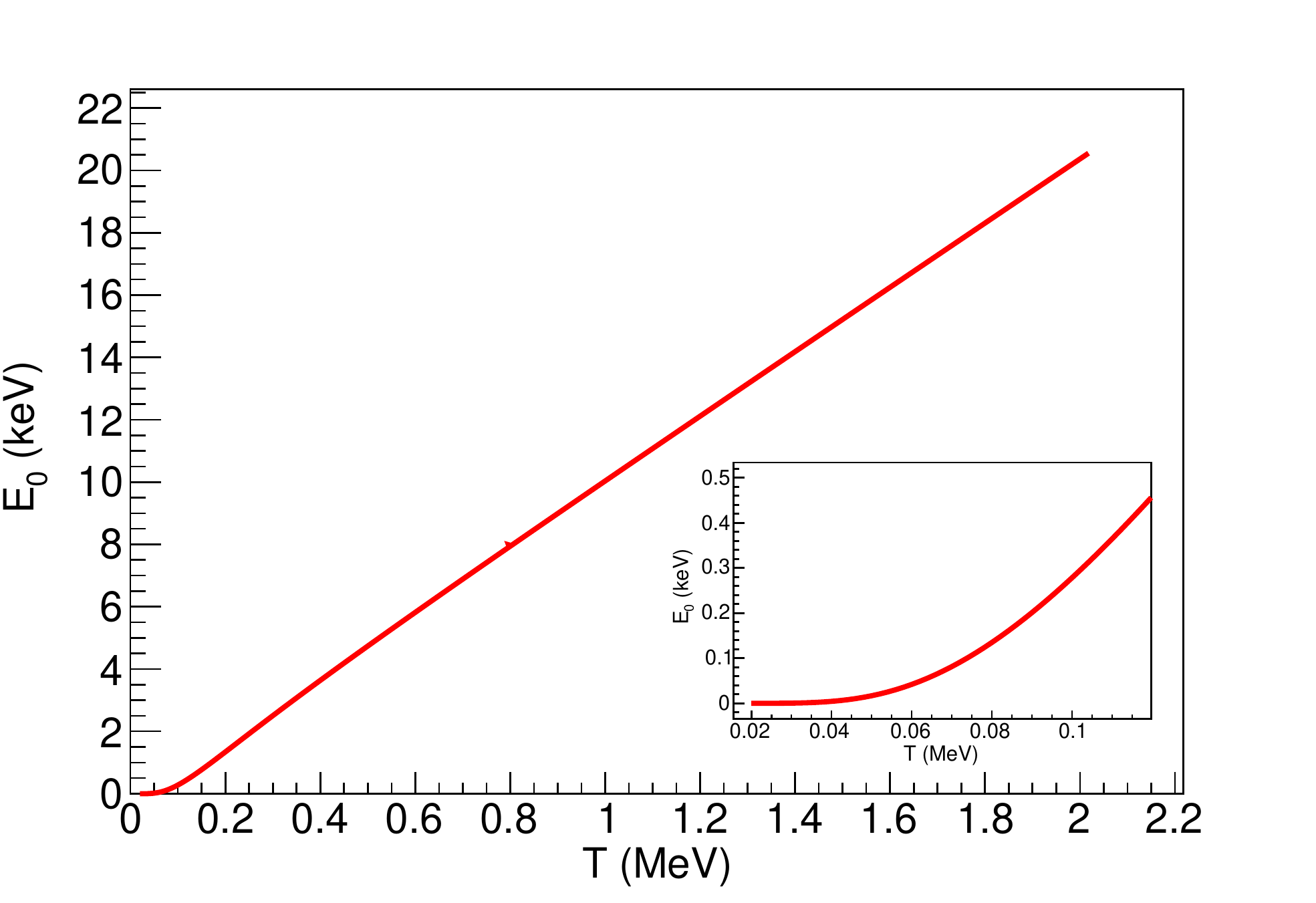}
	\caption{\label{shift_v_T}The energy shift E$_0$ due to relativistic screening as a function of temperature for $Z_1$=2, $Z_2$=4, and $\mu$=0.  The high temperature limit is as given in Equation \ref{high_t}.  The inset 
		shows the shifts at low temperature.}
\end{figure}

This shift is small, $\sim20$ keV for $Z_1 =2$ and $Z_2=4$ at $T\sim2$ MeV. The values of $\Delta$E as a function
of temperature are shown in 
Figure \ref{shift_v_T}; the trend is nearly linear except at low temperature.
Its effect on non-resonant TRRs is also expected to be small, particularly as the cross-section is negligible near $E=0$.
However, the shift could be significant for resonant rates, particularly those low-lying resonances near the threshold,
where even a small shift in the energy can result in a significant change in the cross-section.

Another possible effect of this shift is the influence on sub-threshold resonances.  A shift to higher energy can
result in a much more significant decrease in the cross-section and total reaction rate as less of the sub-threshold
resonance is integrated over.  Using this shift, a possible reduction in the TRR for low-lying resonances is investigated.  The result may be
significant because the energy shift results in less of the resonance tail being included
in the TRR. 
From Figure \ref{shift_v_T}, one expects the enhancement to be small, approximately 1\%, and to scale with temperature.  
This scaling is because as the temperature decreases the Debye screening
length increases, resulting in a reduced energy shift, which approaches zero. This is important to note as the effect is most pronounced only at high temperatures
(early in BBN) and in regions where a significant portion of the resonance may be removed from
the reaction rate - near the threshold. This may be advantageous as there may be a slight reduction in BBN reaction rates during the early stages, resulting in a slower progression to the $A=7$ nuclei.

\begin{figure}
	\includegraphics[width=\linewidth]{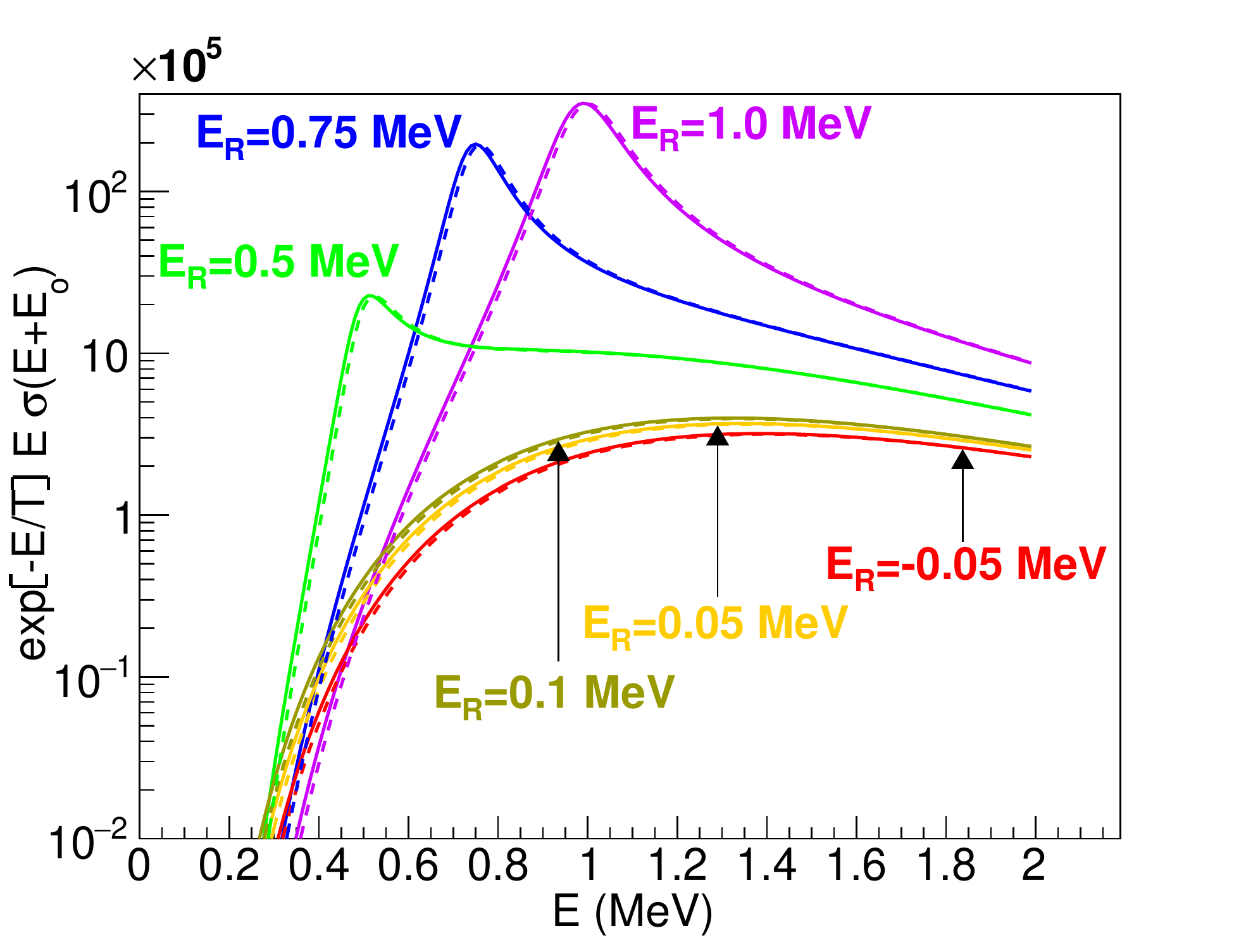}
	\caption{\label{res_int}The integrand of the reaction rate for a narrow resonance for
		three values of the resonance energies for the bare potential (dotted lines) and the screened potential
		(solid lines).  For the low-lying resonances, the
		resonance at 0.1 MeV corresponds to the upper-most set of lines, while the resonance at -0.05 MeV corresponds to the lowest set of lines.}
\end{figure}

The effect of this shift is shown in Figure \ref{res_int} which shows a resonance 
($\Gamma_{x}$ = 110 keV, $T=1$ MeV) in the $^7$Be $+^3$He reaction times
the Boltzmann distribution for six values of $E_R$. This quantity is the integrand of equation 
\ref{a9}.  For this quantity, the particle energy is dictated by the Boltzmann distribution
while the value of the cross section is determined by the energy shift (to higher energy), effectively
shifting the cross-section to lower energy in the particle distribution.  We note, of course, that the 
	lowest resonances in this figure have been eliminated by
	experiment, but they are shown here to indicate the relative strengths of the resonance tails and to emphasize the point that the actual resonance peaks are not 
	important as the penetrability in the entrance channel
	at the lowest energies reduces the cross-section to
	negligible values.

For all resonances shown, the
shift to higher energy shifts the entire integrand towards the
 high-energy tail of the resonance. For the low-lying
  resonances, $E_R\lesssim500$ keV, the integral is increased slightly as the tail of the integrand is emphasized.
For a higher-energy resonance,
  $E_R\gtrsim500$ keV, only a very small portion of the lower
   energy tail is cut out of the integration, as most of the
    low-energy tail is dominated by the entrance channel
     penetrability.

\begin{figure}
	\includegraphics[width=\linewidth]{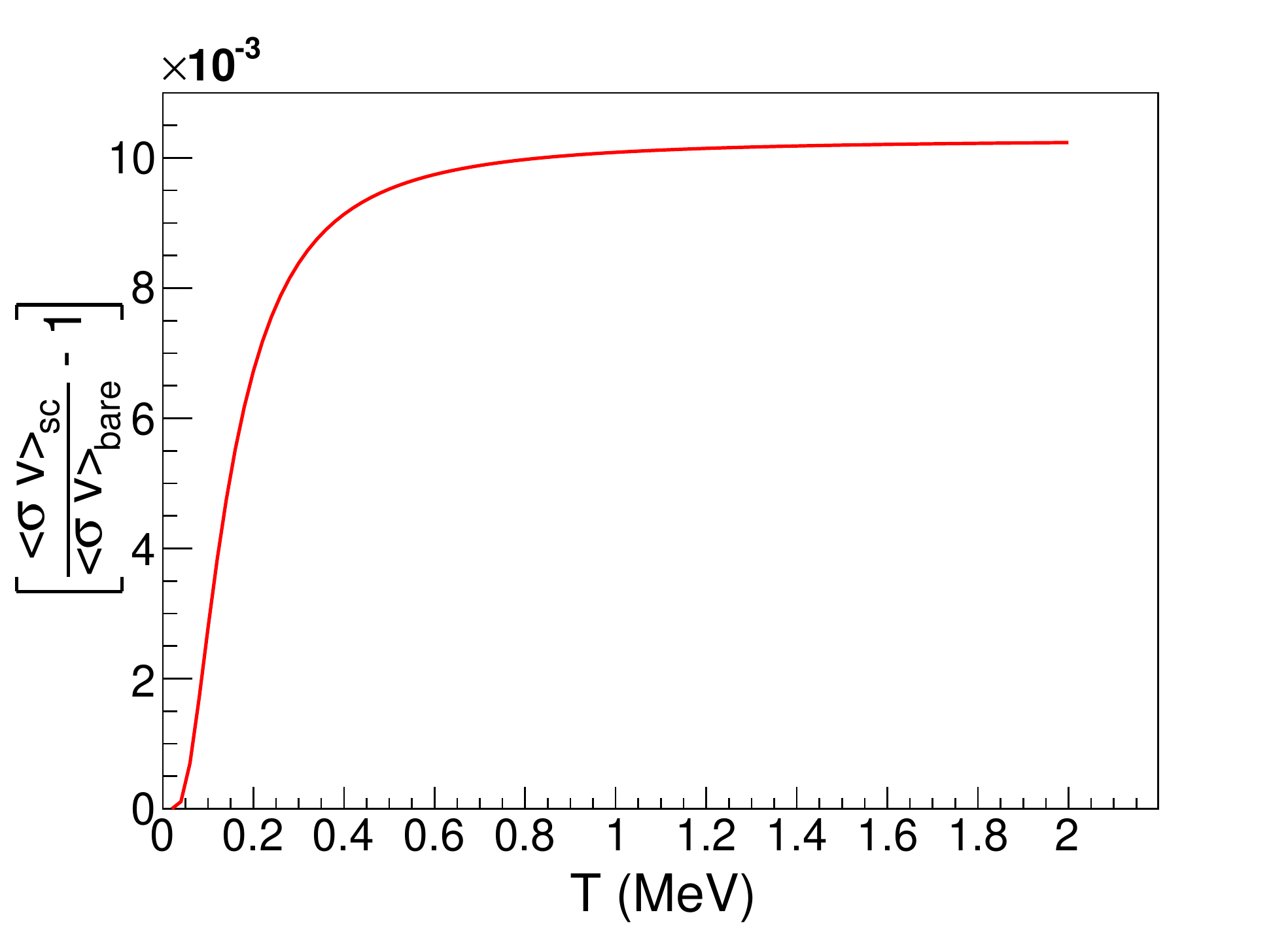}
	\caption{\label{non_res_enh}Ratio of screened to bare TRR for the non-resonant component of the TRR only.}
\end{figure}

The enhancement for non-resonant rates is exemplified in Figure \ref{non_res_enh} for $Z_1=2$ and 
$Z_2=4$.  As expected, the enhancement is small and 
always greater than unity at non-zero temperature since
the cross-section is always monotonically increasing with energy and a small shift to
positive energy results in a larger cross section integrated into the reaction rate.
It was found that this enhancement varies little with resonance energy and width.  This makes sense 
considering the integrand of the reaction rate and the very 
small energy shift $\Delta E$. 
In any case, essentially the entire cross-section is
integrated over, but it is effectively shifted to a higher
energy in the Maxwell-Bolzmann distribution by an amount
$\Delta E$.  For a very small shift, the ratio of rates in 
Figure \ref{non_res_enh} is roughly:
\begin{equation}
    \frac{R_{sc}}{R_{bare}} \sim e^{\Delta E/T}
\end{equation}
which is the Salpeter factor.

\subsection{Accuracy of the Salpeter Approximation}
In the previous section, the exact integration of Equation \ref{a9} was used to
determine the correction to reaction
rates for the relativistic electron gas.  The typical prescription is to use the
Salpeter approximation to 
determine the width corrections.  This approximation results from the separation of
the exponent containing the Debye
length from the integration over energy and use as an independent coefficient in the
reaction rate without shifting
the cross section energy.  

\begin{figure}
	\includegraphics[width=0.9\linewidth]{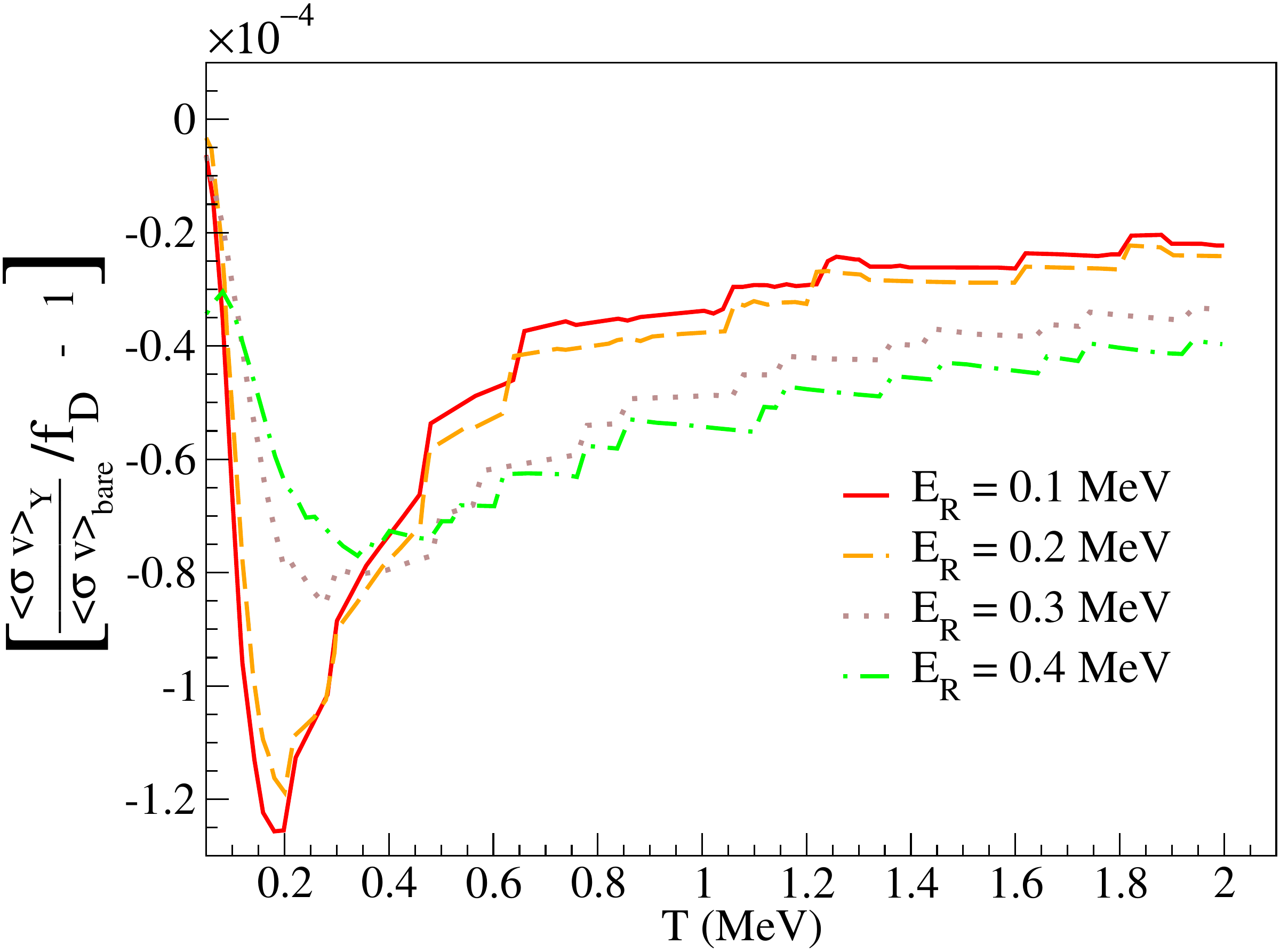}
        \caption{\label{salpeter_compare}The correction to the reaction rate from the
energy shifts as studied in this 
                work divided by the Salpeter correction factor for several resonance energies, indicated
in the figure.}
\end{figure}

This correction can be evaluated by considering the change in reaction rates using
the exact screening due to the Coulomb potential in the Yukawa form compared with the
Salpeter approximation.  This evaluation is shown in Figure \ref{salpeter_compare} where we plot 
the quantity 
\begin{equation}
\label{zz}
      \left.  \left(\frac{<\sigma v>_{\rm Yukawa}}{<\sigma v>_{\rm bare}}\right) \right/ f_D .
\end{equation}
In Figure \ref{salpeter_compare}, we take $\Gamma_\gamma$ = 110 keV, $\Gamma_p$ = 2 keV corresponding to $E_R\approx$ 500 keV, and the difference between the ratio in Equation (\ref{zz}) and unity is multiplied by 10$^4$.  It can be seen
that the relative difference between the
correction in Figure \ref{salpeter_compare} and the Salpeter correction factor is on the order of
10$^{-4}$. The small energy shifts from
electron screening induce a difference ratio nearly equal to the Salpeter
correction factor to within $\approx 0.01$\%, 
though
it  does  appear  that  the  difference  gets  larger  at  lower
temperatures  and  lower  resonance  energies  as  more  of
the resonance falls below the reaction threshold for the
energy shifts induced by screening.
\subsection{Thermonuclear Reaction Rates (TRRs) for Various Resonances}
\begin{figure}
        \includegraphics[width=\linewidth]{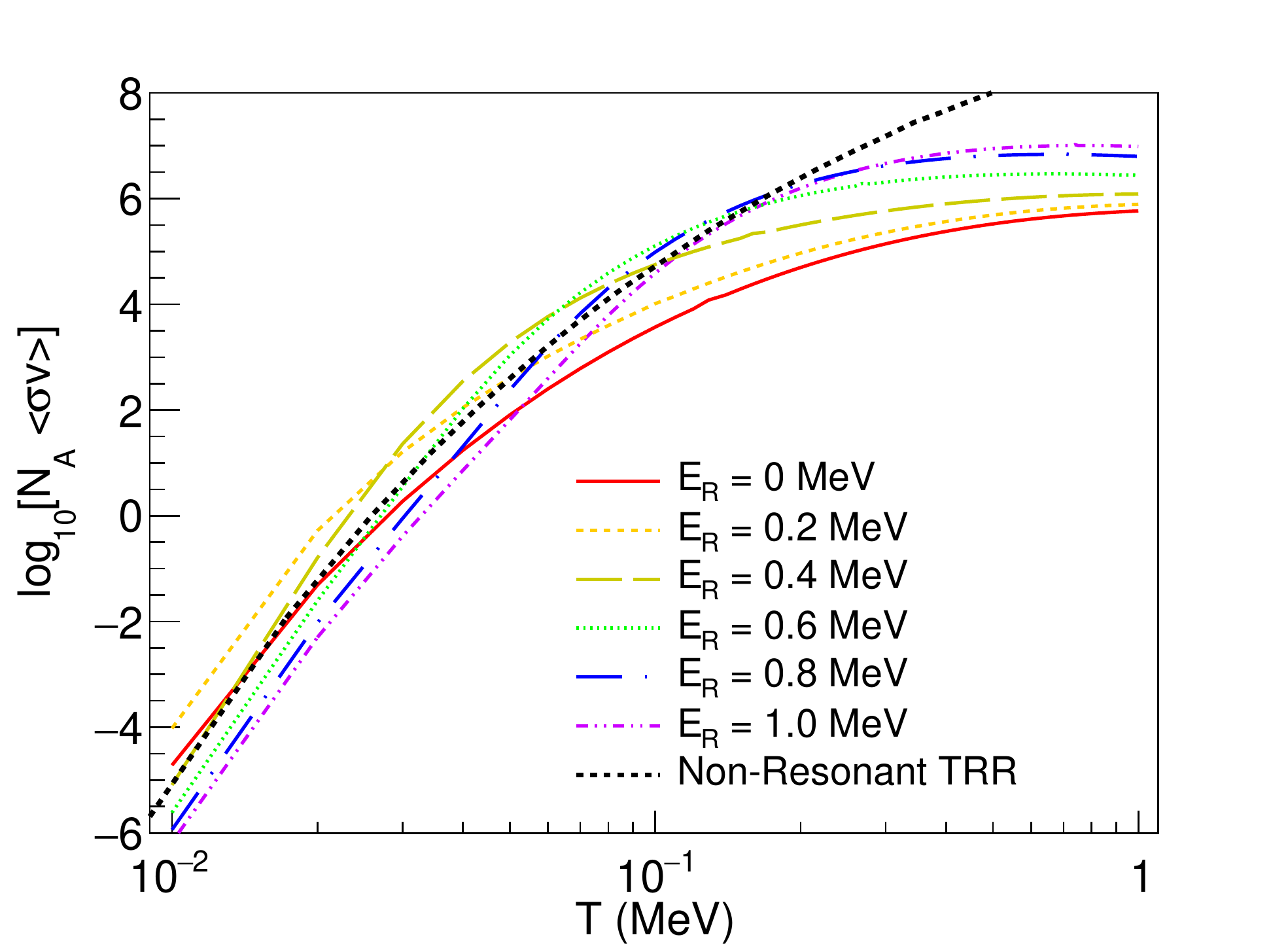}
        \caption{\label{e_dep_wid_rates}Thermonuclear reaction rates for the
$^7$Be($^3$He,$\gamma$)$^{10}$C reaction for
                several resonances for possible particle-decay channels in the $^{10}$C compound nucleus. 
                The dashed black line is the non-resonant TRR.
                Assumptions are described in the text.}
\end{figure}

TRRs for this reaction are shown in Figure \ref{e_dep_wid_rates}
for several resonances. For these resonances, we assume a total $^{10}$C decay width $\Gamma_x=110$ keV, a spin degeneracy
factor $\omega=0.5$, and a 
single particle width $\theta_L^2=0.5$. The thermonuclear rates are shown only for
the resonances and must be
added to the non-resonant rate, also shown in the figure.  We observe that the results in Figure \ref{e_dep_wid_rates} decrease with the decay width.  For decay widths $\Gamma_x\lesssim$ 50 keV, the resonant rates are less than the non-resonant rate. 

It is seen that the rates are similar for all the resonances in this temperature
range.  Several factors 
must be considered.  First, for the energy-dependent widths, a lower resonance may
have a larger particle population in the
Maxwell-Boltzmann distribution, but the entrance channel widths are also smaller.  Very roughly, the
rate is proportional to the 
penetrability factor times the Maxwell-Boltzmann factor.  While the penetrability
factor increases with energy, the 
Maxwell-Boltzmann factor decreases.  

From the Figure \ref{e_dep_wid_rates} one can conclude that resonances in the energy range of 0$\le$E$\le$1 MeV and 
with $\Gamma_x < 110$ keV for the $^7$Be+$^3$He reaction are insufficient for 
reducing the BBN production of $^7$Be. This is consistent with prior results \cite{Broggini12}, and the additional inclusion of relativistic plasma screening effects have also been found to be
negligible.  Thus, the validity of the previous analysis is
maintained.  A possible BBN scenario with a higher density at lower
temperatures may result in such a situation, though one must gauge the effects of the density increase on other reactions as well.
\section{Conclusions}
In this work we explored in detail the consequences of the screening due to the relativistic electron-positron plasma on non-resonant and possible resonances on the secondary reactions destroying $A=7$ nuclei during the Big Bang Nucleosynthesis. We found that effects of screening
from the relativistic plasma are small even for the reaction with the largest $Z_1Z_2$, namely $^3$He+$^7$Be. We note that this reaction remains to be the least experimentally
explored one in the network of BBN reactions.

We scanned through possible resonance parameters (widths and resonance energies) in our calculations. The very small entrance channel widths in any possible resonance renders its effects quite small.  
	BBN reactions would have to overcome this either by resonances to high spin states - which is very unlikely, resonances via neutron captures to destroy A=7 nuclei - which are inhibited by an insufficient neutron abundance by the time A=7 nuclei are produced, or via
	resonances to energy states high above threshold.  As the temperature by the time A=7 nuclei are produced in significant abundance
	is low ($T_9\sim$ 1), this last case is also not highly probable. 

Even though the effects we find are small, it still is worthwhile to demonstrate how robust our current understanding of the BBN is to effects not previously considered. 
This is especially important since the instruments scheduled to go online in the future, such as the Thirty Meter Telescope
 \cite{Skidmore:2015lga}, will measure the abundances of the light
  elements resulting from the BBN with 
  greater precision. 
\vskip 0.2in
\begin{acknowledgments}
MAF and ABB would like to thank the members of of the National Astronomical Observatory of Japan for their hospitality. 
This work was supported in part by Grants-in-Aid for Scientific Research of JSPS (26105517, 24340060) of the Ministry of Education, Culture, Sports, Science and Technology of Japan, in part by the US National Science 
Foundation Grants No. PHY-1204486 at Western Michigan University (WMU)  
and PHY-1514695 at the University of 
Wisconsin, in part by a WMU Support for Faculty Scholars Award and in part by the University of Wisconsin Research
Committee with funds granted by the Wisconsin Alumni Research
Foundation.
\end{acknowledgments}

\appendix*
\section{Appendix: Screening of the Coulomb potential}
In a plasma the Coulomb potential between two nuclei is screened:
\begin{equation}
\label{a1}
V_C^{\rm scr} = \frac{Z_1Z_2e^2}{r} \exp \left( - \frac{r}{\lambda_D} \right) 
\end{equation}
where $\lambda_D$ is the Debye radius. The non-relativistic contribution to the Debye radius is given by  
\begin{equation}
\label{a2}
\lambda_D = \left[ \frac{T}{4 \pi e^2 N \left(\sum_i X_i \frac{Z_i^2}{A_i} + \xi \sum_i X_i \frac{Z_i}{A_i} \right)} \right]^{1/2}
\end{equation}
where $N$ is the ion number density, $X_i$ is the mass fraction of nuclei of type $i$, and $\xi$ is a factor that accounts for the electron degeneracy \cite{Salpeter:1954nc}. Eq. (\ref{a2}) is derived using non-relativistic limit, which is appropriate for the nuclei in the Big Bang. This formula was used in the calculations of Ref. \cite{Wang:2010px}. 

Contribution to the Debye length from the relativistic electron-positron plasma can be calculated exactly to all orders from the Schwinger-Dyson equation for the photon propagator \cite{Kapusta:2006pm}. It is given as 
\begin{equation}
\label{a3}
\frac{\pi^2}{\lambda_D^2} = e^2 \frac{\partial}{\partial \mu} 
\int_0^\infty{dpp^2\left[\frac{1}{e^{(E-\mu)/T}+1}-\frac{1}{e^{(E+\mu)/T}+1}\right]},
\end{equation}
where $E=\sqrt{p^2+m_e^2}$ and $\mu$ is the chemical potential. 

The correction to the reaction rates, 
\begin{equation}
\label{a4}
\langle \sigma v \rangle = \frac{1}{\pi m_{12}} \left( \frac{2}{T} \right)^{3/2} 
\int_0^{\infty} e^{-E/T} E \sigma (E) dE ,
\end{equation}
due to the plasma effects was first calculated by Salpeter \cite{Salpeter:1954nc}. He found that the rates are enhanced by a factor of 
\begin{equation}
\label{a5}
f_D = \exp \left( \frac{Z_1Z_2e^2}{T\lambda_D} \right) .
\end{equation}
A comparison of different derivations of the Salpeter's plasma correction is given in Ref.  \cite{Bahcall:2000bh}. Here we outline another derivation which illustrates the behavior of enhancement in the presence of resonances. The dynamics of two colliding nuclei in a plasma below the Coulomb barrier is described  by \cite{Balantekin:1997yh} 
\begin{equation}
\label{a6a}
H^{\rm scr} \Psi = E \Psi
\end{equation}
where 
\begin{equation}
\label{a6b}
H^{\rm scr} = T + V_N + V_C^{\rm scr} .
\end{equation}
For $r \ll \lambda_D$, the screened Coulomb potential of Eq. (\ref{a1})  can be expanded as 
\begin{equation}
\label{a7} 
V_C^{\rm scr} \sim V_C^{\rm bare} - \frac{Z_1Z_2 e^2}{\lambda_D} = V_C^{\rm bare} -E_0 .
\end{equation}
In the high temperature limit, Equation \ref{a3} yields:
\begin{equation}
\label{high_t}
E_0 = \frac{Z_1Z_2e^3}{\pi} 
\left[\mu^2 + \frac{\pi^2}{3}T^2\right]^{1/2}
\end{equation}
Inserting Eq. (\ref{a7}) into Eq. (\ref{a6a}) one gets
\begin{equation}
\label{a8}
H^{\rm bare} \Psi = ( E+ E_0) \Psi
\end{equation}
where
\begin{equation}
\label{a8b}
H^{\rm bare} = K + V_N + V_C^{\rm bare} 
\end{equation}
with $K$ being the kinetic energy associated with the relative motion of the nuclei and $V_N$ is the attarctive nuclear potential. 
Hence the effect of the screening is to shift the energy by an amount $E_0$ in the calculations performed using the bare Coulomb potential. As a result the reaction rate takes the form
\begin{equation}
\label{a9}
\langle \sigma v \rangle = \frac{1}{\pi m_{12}} \left( \frac{2}{T} \right)^{3/2} 
\int_0^{\infty} e^{-E/kT} E \sigma (E+E_0) dE .
\end{equation}
After a change of variables $E'=E+E_0$, the rate then takes the form 
\begin{equation}
\label{a10}
\langle \sigma v \rangle \sim \int_{E_0}^{\infty} e^{-(E'-E_0)/T} (E'-E_0) \sigma (E') dE' .
\end{equation}
Since $E_0$ is very small the lower limit of the integral can be extended to zero and the term linear in 
$E_0$ multiplying the cross section can be ignored. In Ref. \cite{Bahcall:1996kh} it was shown that the correction due to these approximations is indeed very small. The $E_0/T$ contribution to the exponential  yields the Salpeter enhancement of Eq. (\ref{a5}). Note that Eq. (\ref{a9}) demonstrates that a shift away or towards the peak energy can appreciably alter the reaction rates in the presence of resonances.

\end{document}